\documentclass[11pt]{asaproc}
\usepackage{amsmath,bbm,amsfonts}
\usepackage{graphicx,psfrag,epsf}
\usepackage{enumerate}
\usepackage{amsthm,amssymb}
\usepackage{caption}
\usepackage{subcaption}
\usepackage{mathtools}
\usepackage[round]{natbib}
\usepackage{tikz}
\usepackage[finalnew]{trackchanges}
\usetikzlibrary{automata,arrows,positioning,calc,shapes}
\tikzstyle{arr}=[-latex, black, line width=0.5pt]
\tikzstyle{doublearr}=[latex-latex, black, line width=0.5pt]
\tikzstyle{input}=[font=\small\sffamily\bfseries]
\tikzstyle{rect}=[rectangle, draw=black, font=\small\sffamily\bfseries, inner sep=9pt]
\usetikzlibrary{external}
\usepackage{url} 
\usepackage{algpseudocode}

\RequirePackage[colorlinks,citecolor=blue,urlcolor=blue]{hyperref}

\newcommand{\bx}{\mathbf{x}}
\newcommand{\by}{\mathbf{y}}

\newcommand{\Poi}{\text{Poisson}}
\newcommand{\Ber}{\text{Bernoulli}}

\newcommand{\dif}{\mathrm{d}}

\usepackage{times}

\title{Simulating cloud-aerosol interactions made by ship emissions}

\author{Lekha Patel\thanks{Statistical Sciences, Sandia National Laboratories, PO Box 5800 MS 1202, Albuquerque, NM 87185} \and Lyndsay Shand$^\ast$}
\begin{document}

\maketitle

\begin{abstract}
Satellite imagery can detect temporary cloud trails or \textit{ship tracks} formed from aerosols emitted from large ships traversing our oceans, a phenomenon that global climate models cannot directly reproduce. Ship tracks are observable examples of marine cloud brightening, a potential solar climate intervention that shows promise in helping combat climate change.    
Whether or not a ship's emission path visibly impacts the clouds above and how long a ship track visibly persists largely depends on the exhaust type and properties of the boundary layer with which it mixes. 
In order to be able to statistically infer the longevity of ship-emitted aerosols and characterize atmospheric conditions under which they form, a first step is to simulate, with mathematical surrogate model rather than an expensive physical model, the path of these cloud-aerosol interactions with parameters that are inferable from imagery. This will allow us to compare when/where we would expect to ship tracks to be visible, independent of atmospheric conditions, with what is actually observed from satellite imagery to be able to infer under what atmospheric conditions do ship tracks form.
In this paper, we will discuss an approach to stochastically simulate the behavior of ship induced aerosols parcels within naturally generated clouds. Our method can use\remove{ both real and simulated} wind fields \add{and potentially relevant atmospheric variables} to determine the approximate movement \add{and behavior }of the \remove{cloud and simulated}\add{cloud-aerosol} tracks, and uses a stochastic differential equation (SDE) to model the persistence behavior of cloud-aerosol paths. This SDE incorporates both a drift and diffusion term which describes the movement of aerosol parcels via wind and their diffusivity through the atmosphere, respectively. \remove{The longevity of each parcel is then considered as a random variable with a shared mean and variance along a single track.} 
\remove{A comparison of different ship velocity fields and atmospheric wind conditions with real satellite data will be presented to highlight the suitability of our simulation method to this problem.}
\add{We successfully demonstrate our proposed approach with an example using simulated wind fields and ship paths.}
\end{abstract}

\newpage
\section{Introduction} 
\label{sec:introduction}
\begin{figure}[!h]
		\centering
		\begin{subfigure}{.5\textwidth}
		  \centering
		  \includegraphics[width=\linewidth]{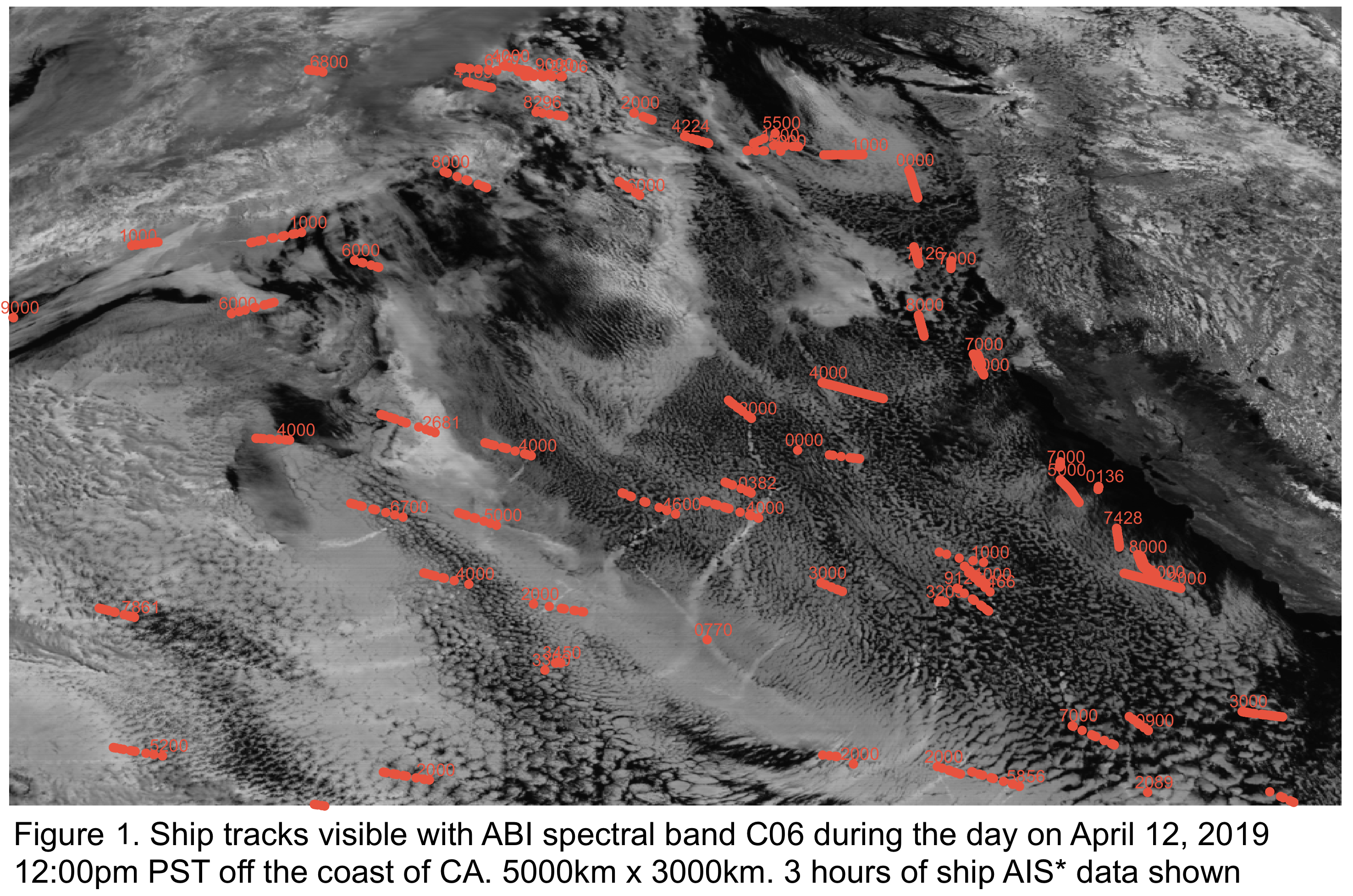}
		  \caption{}
		  \label{fig:sub1}
		\end{subfigure}%
		\begin{subfigure}{.5\textwidth}
		  \centering
		  \includegraphics[width=\linewidth]{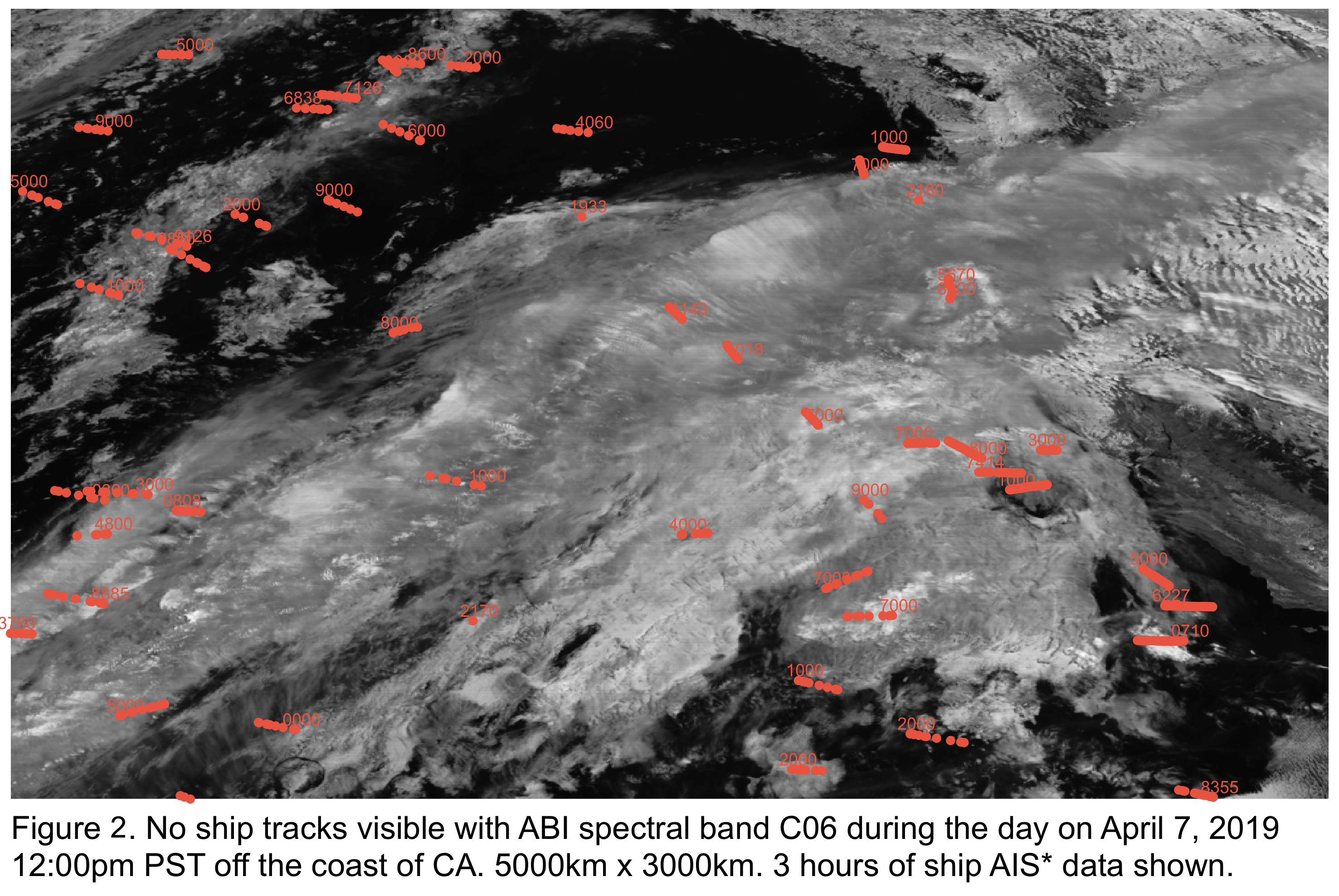}
		  \caption{}
		  \label{fig:sub2}
		\end{subfigure}
		\caption{\footnotesize{Visible ship tracks (left) on April 12, 2019 compared with no visible tracks (right) on April 7, 2019 with 3 hours of known ship locations (shown in red). Images ($5000$km $\times$ $3000$km) taken at 12:00 GMT with ABI spectral band C06 off coast of California.\protect\footnotemark}}
		\label{fig:all}
	\end{figure}

For decades, satellite imagery has been able to detect ship tracks, temporary cloud trails created via cloud seeding by the emitted aerosols of large ships traversing our oceans \remove{, a phenomenon that global climate models cannot directly reproduce}. Ship tracks are of interest because they are unintentional and observable examples of marine cloud brightening, a potential solar climate intervention \citep[e.g.][]{latham1990, ci2015, icc2013}. Ship tracks are visible evidence of the ability of large amounts of \add{anthropogenic} aerosol\add{s} \remove{emissions} to perturb boundary layer clouds enough to alter the albedo of the atmosphere, \add{usually} brightening \remove{and sometimes dimming local}\add{the surrounding} clouds \add{(Twomey Effect, }\cite{twomey1968}\add{)}, and thus significantly contribute to indirect radiative forcing \citep{capaldo99,eyring10}. Recently, this phenomenon has become more frequently observed as \remove{the of }satellite technology has significantly improved since ship tracks were first observed by \cite{conover1966} and \cite{twomey1968}. Using the recently deployed GOES-R \add{geostationary} satellite series, we have seen that these tracks can remain visible in the atmosphere throughout the year,  lasting a few hours to more than 24 hours before diffusing and mixing back into the atmosphere. It is also well-known that certain atmospheric conditions lead to more visible ship tracks than others, varying image sampling times, cloud movement and changes in humidity \citep{possner18} can also cause tracks to be poorly observed and may hinder any corresponding image analysis (see Figure \ref{fig:all}). 

	\footnotetext{GOES-R imagery data available at https://www.bou.class.noaa.gov/saa/products/welcome and AIS data available at https://info.seavision.volpe.dot.gov.}

\add{Although \textit{ship tracks} has been actively studied since the 1960s, indirect radiative forcing is still the largest documented source of uncertainty when it comes to overall radiative forcing in climate modeling} \citep{carslaw2013}\add{. Most current knowledge on specific conditions under which tracks form have come from physical simulation studies under pristine conditions, which do not necessarily represent reality.
In climate simulation studies of this phenomenon} \citep{wang2011, berner2015, possner2018, blossey2018}\add{, aerosol injections are initiated by the user at a known location in fully defined environments. 
Satellite-observed tracks, however, are instead ``initiated" by an unknown source and form in a dynamic and only partially known environment that is difficult or near impossible to replicate in a physical simulation study, which can also be quite computationally expensive. }

\add{In this work, we present a computationally efficient, mathematical simulation approach to emulating the observed formation and behavior of ship tracks.  Exisiting methods focus on modeling the chemical evolution of aerosol composition \citep{riemer08,sofiev09} and are not applicable to the physical modeling of cloud-aerosol paths through the atmosphere. Our approach aims to do so by flexibly accounting for the effect of weather and atmospheric conditions.}

\add{Our method is different from physical simulation approaches in that it attempts to emulate what is observed via satellite, rather than generate the full 3-D micro-physics environment. Ultimately, we would like to infer from imagery and atmospheric data under what conditions do track form or not form to incorporate in our emulation approach. For now, without more information on to what degree atmospheric conditions effect the visibility or behavior, we present the general framework accounting for the cloud movement and point out where atmospheric effects can be incorporated.

The remainder of this paper is organized as follows:} \add{Section} \ref{sec:data} \add{presents the background motivating the simulations. Section }\ref{sec:aerosol_model} \add{and }\ref{sec:imaging_simulation}\add{ outlines our emulation approach and provides simulation examples, respectively. Lastly, Section }\ref{sec:disc}\add{ discusses follow-on work and potential impacts of this research.
 }

%

\section{Background} 
\label{sec:data}
\begin{figure}
\includegraphics[width=0.48\textwidth]{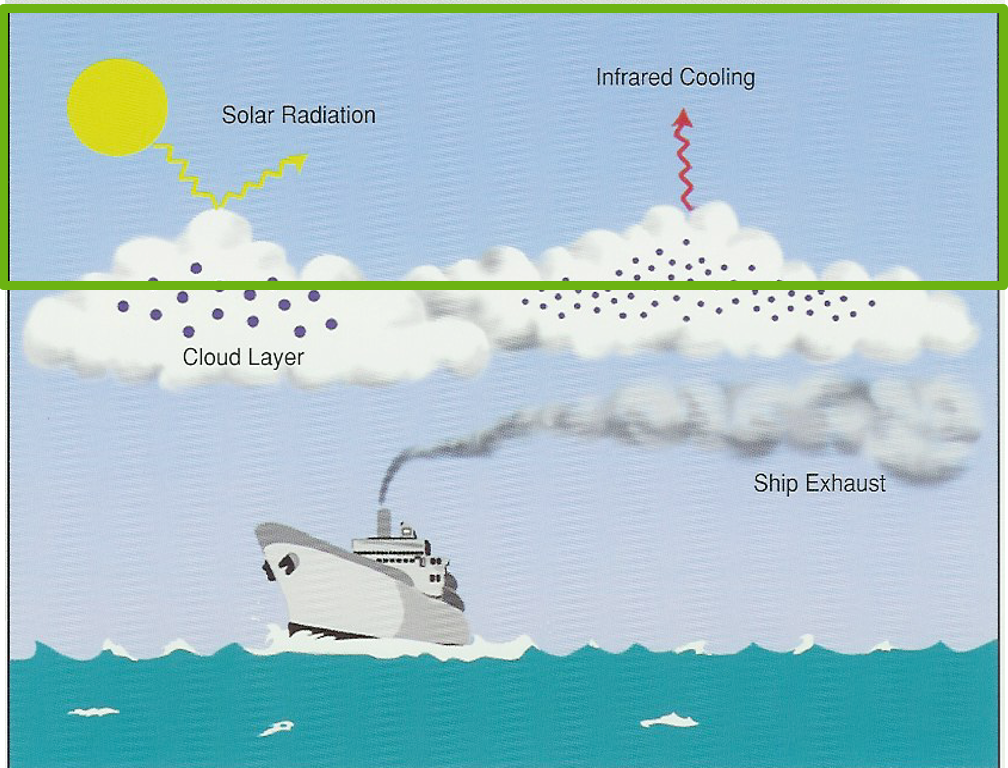}
\includegraphics[width=0.45\textwidth]{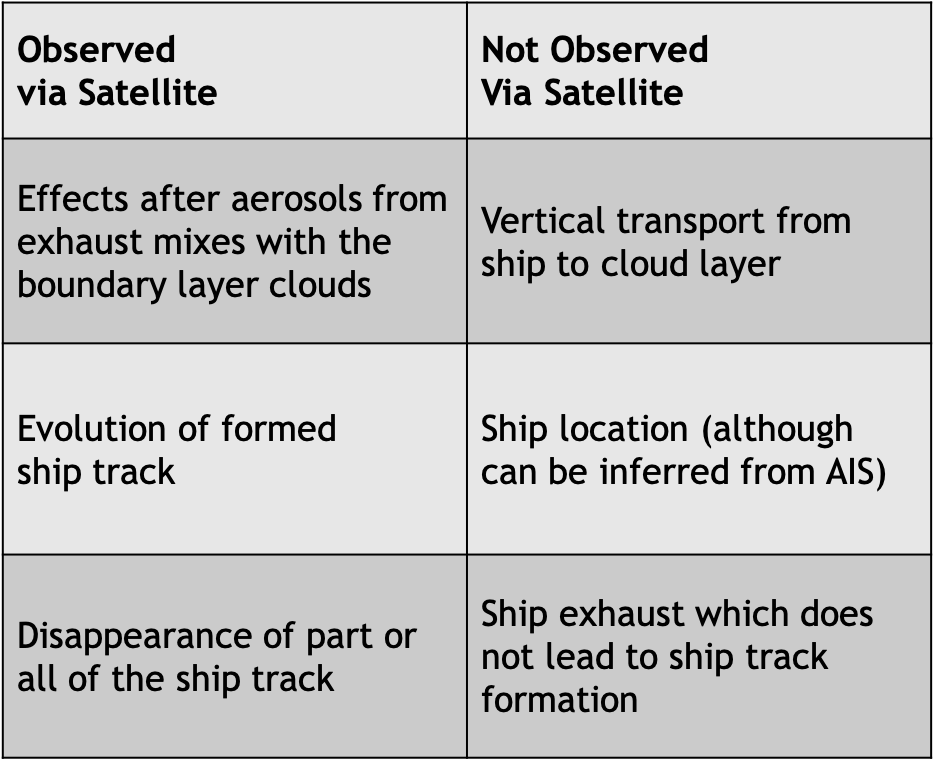}
\caption{\small{https://ral.ucar.edu/staff/jwolff/aerosols.html/intro.html}}
\label{fig:phenom}
\end{figure}

For a given ship, we consider modeling each aerosol \textit{emission burst} as a single target. Each target is transported vertically from the ship through the atmosphere until it reaches a specific altitude near the cloud top height at which the target can become visible to orbital satellites and form a linear tracks in a cloud. 
\add{Figure} \ref{fig:phenom} \add{outlines the general behaviors of the aerosols that are observed or unobserved via satellite.}\add{The green box in Figure} \ref{fig:phenom} \add{represents the portion of the track formation process that is visible via satellite. A ship track is the visible effect of the exhaust aerosols mixing with the low-lying clouds. The vertical transport of the aerosols between the ship's smoke stack and the boundary clouds is largely unknown and unobserved.} 
The exact altitude of the \add{boundary clouds in which the ship track forms} and the time lag between an aerosol burst released from a ship and reaching the visibility height largely depends on the complex weather and cloud dynamics. 
The visibility height can be approximated using cloud top height measurements obtained from satellite retrievals but the time lag is likely impossible to infer from satellite images with spatial resolution greater than a kilometer such as those retrieved from the GOES-R imager. Aerosol transport from ship to boundary layer (height at which cloud formation starts) should be fairly vertical without much resistance but tracking it vertically through the clouds is not trivial. 

 \add{Due to variations in of fuel types and quantities emitted and }\remove{Due to the }complex atmosphere dynamics, not all \textit{emission bursts} will produce visible tracks. \remove{will reach visibility height and thus, with satellite imagery,}\add{Thus,} we only observe ship tracks under the appropriate conditions. This \add{not only} means that not all ship emissions will produce a ship track\add{, but also that}\remove{ and} interruptions in the visibility of an existing ship track can occur\add{ when ships pass under different atmospheric conditions.}
 
\remove{In imagery}\add{To the naked eye}, new ship track observations appear\remove{, for the most part,}\add{in imagery} directly above known ship locations due to the \remove{low }resolution of the imaging so it is reasonable think of the entire vertical transport path from ship to \remove{visibility height }\add{boundary layer}as nearly ``instantaneous" with some epsilon error. For this reason, in this paper, we implicitly impose a known but random time lag between ship emissions and their first detection at the cloud top layer in our simulations. Existing ship track formations will then move with wind dynamics, a variable that is straight-forward to simulate and is independent of actual ship movement.
The visible tracks then persist in the clouds for an unknown time as \textit{ship tracks} until the aerosols are fully diffused into the atmosphere and are no longer distinguishable from the surrounding clouds. 

\section{Modeling aerosols using a Hidden Markov Model (HMM)} 
\label{sec:aerosol_model}
To model the \remove{effects}\add{formation and behavior} of the \remove{cloud }aerosol\remove{s} tracks\remove{, in this section,} we construct a state-space point process representation relating imaging observations of emission tracks to partially observed, known locations of aerosol emission bursts from ships. A constructed Hidden Markov Model (HMM) is outlined \add{section}to characterize the relationship between between the image observations and partially observed truth.
We are interested in building a computational model that can \remove{capture}\add{emulate} the persisting behavior the ship tracks to understand how this behavior changes with changing atmospheric dynamics.

\subsection{State-space representation}
\label{subsec:state_space_model}

The true emission path is generated by the continuously emitted aerosol emission packets by a single ship over the spatial window $\mathcal{X} \subset \mathbb{R}^2$ up to time $ T \in \left[0,\sum_{n=1}^{N-1}\Delta_{n,n+1} \right]$ where $N$ is the number of frames and $\Delta_{n,n+1}>0$ is the time between frames $n$ and $n+1$ (typically between 5 and 15 minutes). For simplicity, we assume in this article that $\Delta_{n,n+1} \equiv \Delta$, so that $t_{n+1}-t_n = \Delta$ for all $n$.

We first define the unobserved spatio-temporal point process $\{X_n: (x,y,t_n) \in \mathbb{R}^2 \times \mathbb{R}\}$ which characterizes the true behavior of the aerosol emission bursts, continuously released prior to (and still visible at) time $t_n$. Second, we define the observed spatio-temporal point process $\{Y_n: (x,y,t) \in \mathbb{R}^2 \times \mathbb{R}\}$ which characterizes the patterns of the partially observed ship tracks in image frame $n$, generated by $X_n$. Using this state-space representation, we formulate a Hidden Markov Model relating the two processes. \add{$T$ can also be defined in terms of number of image frames such that $ T \in \left[0,\sum_{n=1}^{N-1}\Delta_{n,n+1} \right]$ where $N$ is the number of frames and $\Delta_{n,n+1}>0$ is the time between frames $n$ and $n+1$. For simplicity, and since many imagers tend to collect data at regular intervals, we assume that $\Delta_{n,n+1} \equiv \Delta$, so that $t_{n+1}-t_n = \Delta$ for all $n$.}

The true emission path is generated by the continuously emitted aerosol emission packets by a single ship over the spatial window $\mathcal{X} \subset \mathbb{R}^2$ up to \remove{time }time $ T>0, T\in \mathbb{R}$\remove{. }\add{, with $\mathcal{X}$ and time $T$ typically defined by the imager or the user.}
Although in practice, the observed \add{satellite imagery} \remove{images from GOES-R have a temporal resolutions of 5-15 minutes }and our partially observed data $Y_{t_n}$ is observed discreetly, we will treat time as continuous in our simulation model.
For ship $k = 1 \dots K$ which produces a track, we assume that its entire emission path is comprised of $P_k >0$ aerosol bursts\add{(packets)} which may or may not become visible. Assuming that only $k_{t_n}$ of $K$ ships\add{ that are expected to be observed prior to $T$,} have entered the window $\mathcal{X}$ by time $t_n<T$, for an arbitrary single track $k = 1 \dots k_{t_n}$, only $p_{k_{t_n}} \leq P_k$ packets are expected to become visible.
To show proof of concept,  for now we will ignore the complex cloud dynamics and assume all emission packets reach \remove{the cloud top height}\add{reach the boundary layer clouds and become visible} with time lag  $< \epsilon$. This will allow us to start with a general simulation framework and build in more atmospheric conditions when needed at a later time.

In the region of interest $\mathcal{X}$, we denote the set of \underline{true} positions or \textit{states} of each packet as $\{\bx_{i,n}\}_{i=1}^{p_{k_{t_n}}},$ where $\bx_{i,n} \in \mathcal{X}$ denotes the state of the $i$th packet of emission track $k$ at time $t_n$.

Existing ship tracks are only modified at the next time step $t_{n+1}$ in three possible ways:
\begin{itemize}
\item the oldest aerosol emission packets at the end of the track diffuse completely\add{ and mix back} into the atmosphere (leaving no detectable trace), or
\item surviving packets diffuse and become less distinguishable as part of the track (but are still visible), according to cloud dynamics and wind motion, or
 \item new packets appear at the front of the track in the direction the ship movement.
\end{itemize}

These situations result in $p_{k_{n+1}}$ new states\add{(locations)} $\{\bx_{i,t_{n+1}}\}_{i=1}^{p_{k_{n+1}}}$ in each of the new and existing emission tracks present at time $t_{n+1}$.

In \remove{this setup}\add{practice}, however, the full lifespan (from first appearance to permanent disappearance) of each emission packet is \textit{unknown}. Instead, at each observed image frame $n$, the GOES-R ABI sensor captures a snapshot in time of all estimated packet locations without information on age of the packet, i.e. how long the observation\add{s} \remove{has}\add{have} visibly persisted in the atmosphere.
It is also the case that the locations of the emission packets over their lifespan are not unique and can share a location with another emission packet.
Specifically, for a track $k_{t_n}$, a set of $o_{k_{t_n}}\leq p_{k_{t_n}}$ observations $\{\by_{i,t_n}\}_{i=1}^{o_{k_{t_n}}},$ is recorded, where $\by_{i,t_n} \in \mathcal{Y}$ denotes the state of the $i$th observation at time $t_n$. We may assume that $\mathcal{Y} = \mathcal{X}$.

At time $t_n \in \mathbb{R}$, a newly observed track can be generated from newly released emission packets into the atmosphere. Due to the complex dynamics of the atmosphere, it is not often possible to link new observations to their true source. An observed aerosol track from GOES-R is not always visible directly above the known ship location. Thus, we will assume that there is no information about which emission packet generates which observation.
Since there is no ordering on the respective collections of emission packet states and measurements at time $t_n$, they can be naturally represented as finite spatial-temporal point processes. Specifically, for $n = 1, \dots N$, we denote
\begin{align*}
	X_{t_n} &= \{\{\underbrace{\bx_{1,t_n}, \dots, \bx_{p_{1,t_n}}}_\text{\parbox{2.5cm}{\centering $p_{1}$ packets\\[-6pt] from emission $1$}} \}, \dots, \{\underbrace{\bx_{k_{t_n},t_n}, \dots, \bx_{p_{k_{t_n},t_n}}}_\text{\parbox{2.5cm}{\centering $p_{k_{t_n}}$ packets\\[-6pt] from emission $k_{t_n}$}}\}\} \in \mathcal{F}(\mathcal{X}) \quad k_{t_n} \leq K\\
	Y_{t_n} &= \{\{\underbrace{\by_{1,t_n}, \dots, \by_{o_{1,t_n}}}_\text{\parbox{2.5cm}{\centering $o_{1}$ packets\\[-6pt] from emission $1$}} \}, \dots, \{\underbrace{\by_{k_{t_n},t_n}, \dots, \by_{o_{k_{t_n},t_n}}}_\text{\parbox{2.5cm}{\centering $o_{k_{t_n}}$ packets\\[-6pt] from emission $k_{t_n}$}}\}\} \in \mathcal{F}(\mathcal{Y}) \quad 0 \leq o_n \leq p_{k_{t_n}}
\end{align*}
where $\mathcal{F}(\mathcal{X})$ and $\mathcal{F}(\mathcal{Y})$ denote the collections of all finite subsets of $\mathcal{X}$ and $\mathcal{Y}$ respectively. The target point process $X_{t_n}$ is referred to as the \textit{multi-target state} and the measurement set $Y_{t_n}$ is referred to as the \textit{multi-target observation}.
With this model specification, the objective is to recover the true states of emission packet point processes $X_{t_1}, X_{t_2}, \dots, X_{t_N}$ from their measurement sets $Y_{t_1}, Y_{t_2}, \dots, Y_{t_N}$.

\subsubsection{Multi-target state model} 
\label{ssub:multi-target_state_model}
In this section, we describe a finite point process model for the time evolution of the multiple-target state \add{$X_{t_n}, n=1,\ldots,N$}, which incorporates emission packet motion, birth and death. Specifically, we mathematically define the processes of aerosol packets first being conceived in \remove{the cloud-top layer}\add{boundary layer clouds}, their motion and diffusion through the atmosphere until their permanent \remove{unobservance}\add{disappearance}.

After an aerosol track has already formed at time $t_{n-1}$, if an emission packet $\bx_{t_{n-1}} \in X_{t_{n-1}}$ \remove{of it }\add{which makes up part of that track} survives to time $t_{n+1} > t_n$, its subsequent state is determined by a \textit{drift} term which is described by the wind motion at $\bx_{t_{n-1}}$, and a \textit{diffusion} term which describes the diffusion of the emission packet within the clouds it is situated in. This type of process is known as a \textbf{Markov diffusion process} and is described by the following (continuous time) stochastic differential equation:
\begin{align}
	d \bx_t = \underbrace{\mu(\bx_t,t)}_{\text{drift}} dt + \underbrace{\sigma(\bx_t,t)}_{\text{diffusion}} dB_t, \label{markov_diffusion_eq}
\end{align}
where $B_t \sim \mathcal{N}_2(\mathbf{0},tI_2)$ denotes a standard Brownian motion in two dimensions, with $I_2$ denoting the 2-dimensional identity matrix. The drift function $\mu(\bx_t)$ denotes the \textbf{wind velocity} at point $\bx_t$, at time $t$ and is in general known. For this problem, we choose the diffusion function $\sigma(\bx_t) \equiv \sigma_x$ to be a \textbf{constant} that describes the diffusivity of an aerosol parcel within the atmospheric boundary layer. The solution to (\ref{markov_diffusion_eq}) with a changing wind velocity in space and time is in general unknown and requires numerical solvers which may be computationally cumbersome and time consuming. For simulation purposes, we therefore propose the following approximation.

Given discrete time intervals of the form $I_n = (t_n, t_{n+1}] \equiv (n\Delta, (n+1)\Delta],$ with $n \in \mathbb{Z}^+$, we assume that the simulation interval time $t_{n+1} - t_n = \Delta$ is taken small enough so that the wind velocity within the interval is approximately constant. That is to say, for a continuous time point $t \in I_n$, we use the approximate SDE
\begin{align}
	d\bx_t = \mu(\bx_{t}) dt + \sigma_x dB_{t}, \label{approx_markov_diffusion_eq}
\end{align}
with $\mu(\bx_{t})$ denoting the wind velocity field for a parcel with state $\bx_{t}$\remove{ at time $t_n = n\Delta$}. Given \remove{its}\add{a previous} state $\bx_s$ at time $s \in I_n, t>s$, equation (\ref{markov_diffusion_eq}) can be solved directly
\begin{align*}
	d \bx_t & = \mu(\bx_{t}) dt + \sigma_x dB_t \\
	\implies \bx_t - \bx_s & = \int^t_s \mu(\bx_{t}) \; \dif w + \sigma_x (B_t-B_s) \\
	& = \mu(\bx_{t}) (t-s) + \sigma_x B_{t-s},
\end{align*}
where $B_t-B_s \stackrel{D}{\equiv}\footnote{$\stackrel{D}{\equiv}$ denotes equivalence in distribution.} B_{t-s} \sim \mathcal{N}_2(\mathbf{0},(t-s)I_2)$. This implies the corresponding transition density is
$$\bx_t|\bx_s \sim \mathcal{N}_2(\bx_s +  \mu(\bx_{s}) (t-s),\sigma_x^2(t-s)I_2).$$
In particular, the probability density of the parcel's transition to state $\bx_{t_n} \in X_{t_n}$ from state $\bx_{t_{n-1}}$ is given by the Markovian density $f_{t_n|t_{n-1}}^M(\bx_{t_n}|\bx_{t_{n-1}}) \sim \mathcal{N}_2(\bx_{t_{n-1}}+\mu(\bx_{t_{n-1}})\Delta,\sigma_x^2 \Delta I_2)$. Its behavior at this time is therefore modeled by the point process $S_{t_n|t_{n-1}}(\bx_{t_{n-1}})$, where
\begin{equation} \label{def:PP of surviving targets}
	S_{t_n|t_{n-1}}(\bx_{t_{n-1}}) = \begin{cases}
\bx_{t_n} & \text{where} \; \bx_{t_n} \sim f_{t_n|t_{n-1}}^M(\cdot|\bx_{t_{n-1}}) \; \text{with probability} \; p_{S,t_n}(b_{\bx_{t_{n-1}}}) \\
\emptyset & \text{otherwise}.
\end{cases}
\end{equation}
Here, $p_{S,t_n}(b_{\bx_{t_{n-1}}})$ denotes the \textit{survival probability} of packet $\bx_{t_{n-1}}$at time $t_n$ (described in more detail below) and where the motion diffusion coefficient $\sigma_x$ is unknown and requires estimation from the model.

A new emission packet at time $t_n \in \mathbb{R}$ can arise in two ways. The first is as a spontaneous birth (of a newly risen emission track), which is independent of any existing emission track. The second is by \textit{spawning} from an existing emission source, resulting in a newly visible emission packet. We denote the \textit{birth time} of packet $\bx_{t_n}$ (partially) observed at time $t_n$ as $b_{\bx_{t_n}}$.

Spontaneous births of new emission tracks at time $t$ are denoted by the finite point process $\Gamma_t$. We model $\Gamma_t$ as a finite \textbf{Poisson point process} with \textit{intensity function} $\gamma_t(\bx) = \lambda_{\gamma_t} f_{b,t}(\bx))$, for $\mathbf{x} \in \mathcal{X}:$
\begin{equation} \label{def: birth PP}
	\Gamma_t \sim \Poi(\lambda_{\gamma_t} f_{b,t}(\bx)).
\end{equation}
\begin{itemize}
	\item Here, $N_{b,t} \sim \Poi(\int_{\mathcal{X}} \lambda_{\gamma_t} f_{b,t}(\bx) \dif \bx)$ denotes the number of births occurring in $\mathcal{X}$ at time $t$.
	\item $f_{b_t}(\mathbf{x})$ denotes their spatial distribution.
\end{itemize}
Assuming we have \remove{simulated }knowledge \add{(simulated or real) }of the boat positions/path that produce these new emissions, we may let this inform $f_{b_t}(\mathbf{x})$. Specifically, if $\mathbf{x}_{b,t_n}$ is the position of a new boat at time $t_n$, then $f_{b_{t_n + \epsilon}}(\mathbf{x}) = \mathcal{N}_2(\mathbf{x}_{b,t_n},\sigma^2_b I_2)$ where \remove{$\phi$}\add{$\epsilon$} denotes the time lag between ship emission and aerosol observation at the cloud\remove{top}\add{boundary} layer.

Spawned births occurring within the time \remove{$t \in (t_{n-1}, t_n]$}\add{interval $I_{n-1}$} denote newly visible emission packets from existing emission tracks that reach the cloud top layer at time $t_n$. Newly spawned targets can only be spawned by packets that were birthed in the previous time interval \remove{$(t_{n-2}, t_{n-1}]$}\add{$I_{n-2}$}, as this models the continuous emission of aerosol packets in a single stream.

\remove{To deal with this uncertainty, w}\add{W}e model the set of spawned births $B_{t_n|t_{n-1}}(\bx_{t_{n-1}})$ at time $t_n$ from a packet $\bx_{t_{n-1}}$ at time $t_{n-1}$ as a finite point process. An example used in this paper is \textbf{Bernoulli point process} with spawning probability $p_{\beta,t_n}$:
$$B_{t_n|t_{n-1}}(\bx_{t_{n-1}}) = \begin{cases}
\{\bx\}; \; \bx \sim f_{t_n|t_{n-1}}^\beta(\bx|\bx_{t_{n-1}}) \; \text{with probability} \; p_{\beta,t_n} & t_{n-2} < b_{\bx_{t_{n-1}}} \leq t_{n-1}\\
\emptyset & \text{otherwise}.
\end{cases}$$
	\begin{enumerate}
		\item The number of spawned targets $N_{s,t_n}$ from $\bx_{t_{n-1}}$ follows $N_{s,t_n} \sim \Ber(p_{\beta,t_n}).$
		\item Therefore \underline{at most one} packet can be spawned from a target $\bx_{t_{n-1}}$ born in the previous time step.
		\item If $N_{s,t_n} = 1$, then the spatial distribution of the spawned target $\bx$ follows $f_{t_n|t_{n-1}}^\beta(\cdot|\bx_{t_{n-1}})$ from $\bx_{t_{n-1}}$.
	\end{enumerate}

In this paper, we assume knowledge of ship positions that continuously emit aerosols whilst moving, thereby corresponding to this spawning process. For simulation purposes, we therefore use the spawning density $$f_{t_n|t_{n-1}}^\beta(\bx|\bx_{t_{n-1}}) = \mathcal{N}_2(\mathbf{x}_{b,t_n - \epsilon} + \epsilon\sigma^2_\beta I_2).$$

The spawning probability $p_{\beta,t_n}$ is directly related to the number of aerosol packets each ship emits during the observed time window. For simulation purposes, we assume that each boat continuously emits aerosols up to the simulation time \remove{$N\Delta$}\add{$T=N\Delta$} and that they exist when within the observed window $\mathcal{X}$. This enables $p_{\beta,t_n} = 1$ when $t_n \leq T$ and is zero otherwise. 

\begin{figure}[!ht]
\centering
\includegraphics[scale=.7]{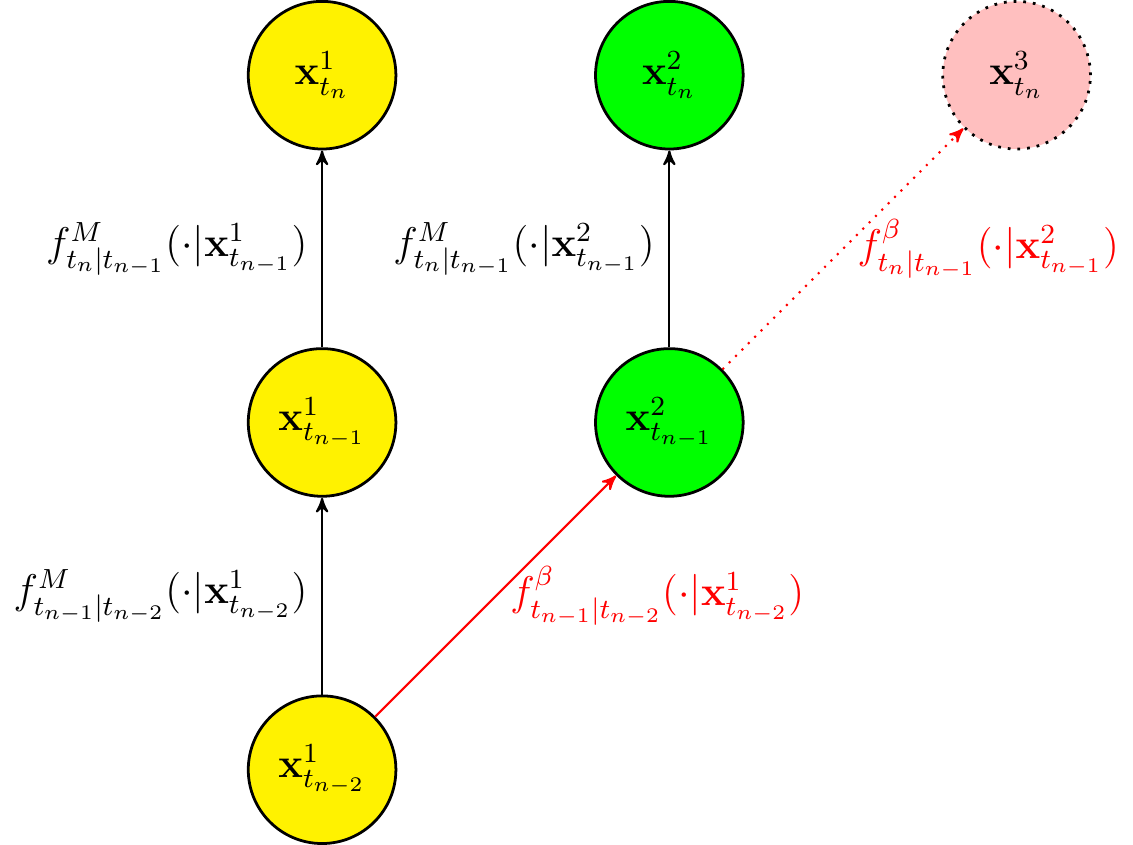} 
\caption{The motion and spawning process. The first packet $\bx_{t_{n-2}}^1$ (indicated in yellow) born at time $t_{n-2}$ undergoes Markovian motion (black) to reach state $\bx_{t_{n-1}}^1$ and spawns a target (indicated in green) $\bx_{t_{n-1}}^2$ at time $t_{n-1}$. This packet undergoes Markovian motion (black) to reach state $\bx_{t_n}^2$ and is to spawn a packet (indicated in pink) $\bx_{t_n}^3$ at time $t_n$.} \label{Figure: spawning process} 
\end{figure}

Figure \ref{Figure: spawning process} illustrates the motion and spawning process of an aerosol packet described by our procedure. 

For a given multi-target state $X_{t_{n-1}}$ at time $t_{n-1}$, each packet $\bx \in X_{t_{n-1}}$ either continues to exist (survives) at time $t_n > t_{n-1}$ with probability $p_{S,t_n}(\bx,b_\bx)$, or ``dies'' with probability $1-p_{S,t_n}(\bx, b_\bx)$. Here, a ``death'' of an emission packet occurs when it sinks back through the atmosphere and ceases to be visible. Furthermore, the survival probability of each emission packet is written as a function of the time $t_n$, its spatial location $\bx$, and the packet's ``birth'' time $b_\bx$ in the region. However, since the effects of up and downward drafts in the atmosphere on each packet can be considered negligible, this enables the survival probability to only be a function of $t$ and $b_\bx$, i.e. that $p_{S,t}(\bx,b_\bx) \equiv p_{S,t}(b_\bx)$.

In this simulation model, we assume that each boat produces a cloud-aeorosol track that has an \textit{average} lifetime $T_d \sim \text{Exp}(\lambda_T)$ from birth. Given $T_d$, the individual aerosol packets that are contained in its emission then each have an \add{independently and identically distributed (i.i.d.)}\remove{iid} death time $$d \sim \text{Log-normal} \left( \mu_d = \log \left( \frac{T_d}{\sqrt{\sigma_{p_d}^2 + T_d^2}} \right), \sigma_d^2 = \log \left( \frac{\sigma_{p_d}^2 + T_d^2}{T_d^2} \right) \right),$$ where $\sigma_{p_d}^2$ is the variance of the packet death time and is a fixed simulation input.

Altogether, we have a multi-target state $X_{t_n}$ of the following form
\begin{equation} \label{def: X_{t_n}}
	X_{t_n} = \underbrace{\left[\bigcup_{\bx \in X_{t_{n-1}}} S_{t_n|t_{n-1}}(\bx)\right]}_{\text{Survived packets}} \cup \underbrace{\left[\bigcup_{\bx \in X_{t_{n-1}}} B_{t_n|t_{n-1}}(\bx)\right]}_{\text{Spawned packets}} \cup \underbrace{\Gamma_{t_n}}_{\text{New emissions}}.
\end{equation}
An important modeling characteristic is that the unions in (\ref{def: X_{t_n}}) are \textit{independent}.

\subsubsection{Multi-target observational model} 
\label{ssub:multi-target_observational_model}
In this section, we describe a finite point process model for the time evolution of the multiple-target observation,\add{$Y_{t_n},n=1,\ldots,N$,} which incorporates observations generated from emission tracks\add{.}\remove{ and those falsely generated by changes in cloud or atmospheric humidity.}

When a packet $\bx_{t_n} \in X_{t_n}$ is \remove{detected}\add{generated according to the process described above}, an observation of it $\by_{t_n} \in Y_{t_n}$ is generated from an observational model $g_{t_n}(\cdot|\bx_{t_n})$. This function is typically chosen to take the form
$\by_{t_n}|\bx_{t_n} \sim \mathcal{N}_2(\bx_{t_n},\Sigma_{\bx_{t_n}}),$ where $\Sigma_{\bx_{t_n}}$ can be taken to be the \textit{marginal} covariance of $\bx_{t_n}$. Specifically, for packet $\bx_{t_n}$ birthed at time $b_{\bx_{t_n}}$, its marginal density can be calculated via
\begin{align*}
	f(\bx_{t_n}) &= \int_\mathcal{X} f^M_{t_n|b_{\bx_{t_n}}}(\bx_{t_n}|\bx_{b_{\bx_{t_n}}}) \pi(\bx_{b_{\bx_{t_n}}}) \; \dif \bx_{b_{\bx_{t_n}}},
\end{align*}
with $\pi(\bx_{b_{\bx_{t_n}}})$ being the initial probability density of packet $\bx_{t_n}$ in $\mathcal{X}$ at the time of its birth. For this paper, we take $\pi(\bx_{b_{\bx_{t_n}}}) = \delta_{\bx_{b_{\bx_{t_n}}}}(\bx_{b_{\bx_{t_n}}})$, the dirac delta function centered at $\bx_{b_{\bx_{t_n}}}$, yielding
$\Sigma_{\bx_{t_n}} = \sigma^2_x (t_n-\bx_{b_{\bx_{t_n}}}) I_2$ and
$$ \by_{t_n}|\bx_{t_n} \sim \mathcal{N}_2(\bx_{t_n}, \sigma^2_x (t_n-\bx_{b_{\bx_{t_n}}}) I_2).$$

When simulating across pixelated grids, we discretize the above equation such that the pixel intensity of a pixel $P$ at time $t_n$ denoted $\mathcal{I}_{t_n}(P)$ follows $$\mathcal{I}_{t_n}(P) \propto \sum_{\by \in Y_{t_n}} \int_{P} f(\by|\bx_{t_n}) \; \dif \by$$ with the normalization constant given by the highest pixel intensity simulated across the video.

For observations generated by true emission packets, we note that a packet $\bx \in X_{t}$, at time $t$ is only detected\add{ by satellites} with probability $p_{D,t}(\bx)$. This \textit{detection probability} has a spatio-temporal dependence structure which is needed to first, model the spatial randomness of cloud humidity/density and second, to account for cloud movement across the observation time window. In the field of view $\mathcal{X}$, if the cloud humidity is too low or too high, emission packets cannot be detected. In the former case, packets cannot be observed since clouds cannot form to produce the necessary observations. In the latter, the cloud density may be too high, or may already be contaminated with existing aerosols which would subsequently not produce observations of new packets.

To deal with this, we may choose to model $p_{D,t}(\bx)$ as a function of the existing cloud humidity/density.
This may be formulated by modeling pixel intensities measured by \add{an imager, such as }the GOES-R ABI sensor,
and utilizing a lower and upper threshold $\iota_L, \iota_U$. For example, setting
\begin{equation} \label{def:probability of detection}
p_{D,t_n}(\bx) = \begin{cases}
	1 & \text{if} \; \iota_L < \mathcal{I}_{t_n}(\bx) < \iota_U \\
	0 & \text{otherwise},
\end{cases}
\end{equation}
enables a packet to be observed with probability one if its true location $\bx$ lies within a pixel of the $n$th frame, with an intensity $\mathcal{I}_{t_n}(\bx) \in (\iota_L, \iota_U)$, sufficient for \remove{an observation of it to be generated}\add{it to be observed via satellite}.

Subsequently, the \textbf{observational point process} $\Theta_{t_n}(\bx_{t_n})$ from an emission packet $\bx_{t_n} \in \mathcal{X}$ follows
\begin{equation} \label{def:PP of target generated observations}
	\Theta_{t_n}(\bx_{t_n}) = \begin{cases}
	\{\by\} \; \text{where} \; \by \sim \by_{t_n}|\bx_{t_n} & \text{with probability} \; p_{D,t_n}(\bx_{t_n}) \\
	\emptyset & \text{with probability} \; 1-p_{D,t_n}(\bx_{t_n}).
\end{cases}
\end{equation}

Altogether, we have a multi-target observation $Y_{t_n}$ of the following form
\begin{equation} \label{def: Y_{t_n}}
	Y_{t_n} = \bigcup_{\bx \in X_{t_n}} \Theta_{t_n}(\bx).
\end{equation}

\section{Imaging simulation} 
\label{sec:imaging_simulation}
In this section, we use the aforementioned simulation method to simulate a video of cloud-aerosol tracks with input parameters as outlined in Section \ref{sec:aerosol_model}. 

\begin{figure}[!h]
\centering
\includegraphics[width=\textwidth]{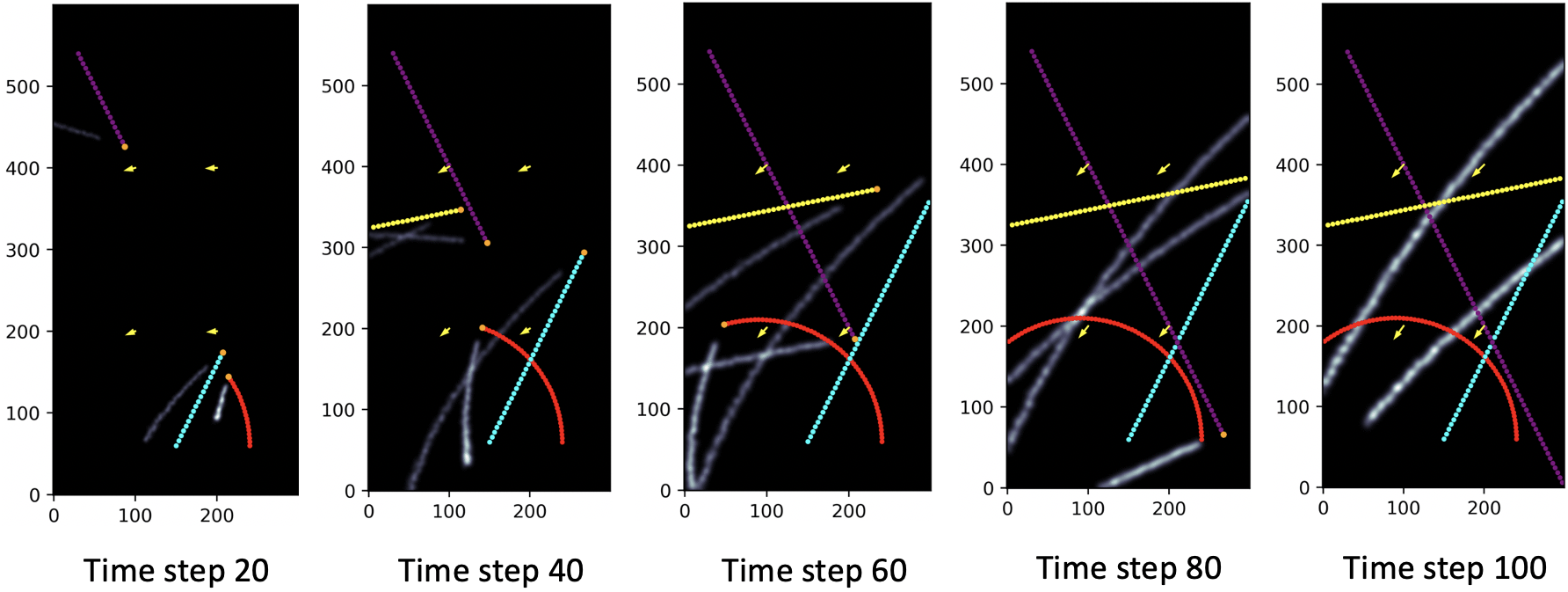}
\caption{\remove{Imaging} Simulation snapshots taken \remove{at 20 time units}\add{4 hours} apart. Boats (red, blue, purple, yellow) and heads (orange) are indicated by colored dotted trajectories, wind direction indicated by the yellow arrows and cloud-aerosol tracks indicated by white trajectories. Parameter inputs for this simulation can be found in Table \ref{table:sim_params}.}
\label{fig:simulation}
\end{figure}

A snapshot of five images taken 20 time steps apart \add{with time step $\Delta=0.2$} hours from the simulation are shown in Figure \ref{fig:simulation}. In this simulation, four boat \add{paths}\remove{(and their paths)} were simulated in longitude/latitude \remove{space units at differing appearance times}\add{appearing at staggered times into the frame}. The\remove{ir} \add{four corresponding} cloud-aerosol tracks were \remove{computed}\add{generated} using \remove{their}\add{the boat} positions, the spawning, persistence and death processes \remove{previously }described \add{in this work with}\remove{and other} input parameters (see Table \ref{table:sim_params}) within a (simulated) circular wind motion. 
A \add{realistic }cloud image was initially used as a background, in which cloud pixels also moved with the simulated wind field. For illustration purposes, however, this has been omitted from the presented images to fully demonstrate the ability of our algorithm in simulating cloud-aerosol tracks. In this simulation, the tracks are observed to \remove{both }follow both the \add{general direction of the boat} path \remove{of its emission boat }and the wind\add{ field}, with the diffusivity effect emphasized by the broadening of each track through time. Furthermore, the pixel intensities are observed to be higher when cloud tracks overlap (as seen at time steps 40, 60 and 80), highlighting the \add{the potential for an} interaction effect of multiple aerosol packets from different emission sources in an arbitrary area. 

\begin{table}[!h]
	\centering
	\begin{tabular}{||c|c||}
		\hline
		Parameter & Value \\
		\hline
		N & 100 \\
		$\Delta$ & 0.2 hours \\
		$\epsilon$ & 5 hours \\
		$\sigma_\beta$ & 0.01\\
		$\sigma_x$ & 0.01 \\
		$\lambda_T$ & 80 hours \\ 
		$\sigma_{pd}$ & 0.2 hours \\ 
		$\mu(\mathbf{x}_t)$ & $\frac{10\pi}{4 N \Delta} \sqrt{(x-0.2t)^2 + (y+0.1t)^2}$ \\
		Position of red boat $\mathbf{x}_{b_1,t}$ & $\left[ 5 \cos\left(\frac{\pi t}{10 N \Delta}\right) +3, 5 \sin\left(\frac{\pi t}{2 N \Delta}\right) + 2 \right]$\\
		Position of blue boat $\mathbf{x}_{b_2,t}$ & $\left[ 1 + \frac{5t}{N\Delta}, 18 - \frac{2t}{N\Delta} \right]$ \\ 
		Position of purple boat $\mathbf{x}_{b_3,t}$ & $\left[1 + \frac{5t}{N\Delta}, 18 - \frac{10t}{N\Delta} \right]$ \\ 
		Position of yellow boat $\mathbf{x}_{b_4,t}$ & $\left[-4 + \frac{10t}{N\Delta}, 10 + \frac{2t}{N\Delta} \right]$ \\ 
		\hline
	\end{tabular}
\caption{Input parameters for simulation.}
\label{table:sim_params}
\end{table}


\section{Discussion and follow-up work} 
\label{sec:disc}
In this paper, we have described a computational method to simulate cloud-aerosol tracks \add{given}\remove{using} wind and boat simulated fields, using a stochastic differential equation that incorporates aerosol packet birth, motion, diffusion and death. A simulation example has been provided to highlight each step of our algorithm. 

Using the presented methodology, a next step would be to verify that this surrogate model is accurate in representing cloud-aerosol paths that are observed in satellite imagery. This is challenging as real cloud-aerosol tracks \add{have an unknown or unidentifiable source and the relationship between observed atmospheric properties and track behavior is not trivial to infer from imagery alone.} \remove{ partially result from some unobservable atmospheric properties that are difficult to replicate in a simulation setting. However, i}\add{I}ncorporating \add{observed wind data from ERA-5 reanalysis and} available atmospheric information that \add{are well-documented to }contribute to cloud track formation such \remove{as information on observed wind fields, }cloud condensation nuclei (CCN) and liquid water paths (LWP) \add{also available from reanalysis products such as MERRA-2}, into an improved simulation algorithm would aid \add{in the simulation of realistic cloud-aerosol behaviors}\remove{analytic image comparisons between the two}. This would thereby allow us to focus on developing methodology and inference mechanisms to estimate \add{atmospheric conditions under which ship tracks form or not form,}\remove{known imaging parameters} which can then \remove{be }directly \remove{compared to }\add{ inform }simulation inputs\remove{ and unknown aerosol properties with particular reference to their longevity in the atmosphere}. \add{Additionally, this level of inference}\remove{This} would require labeling image pixels and extracting observation points of cloud tracks, which may necessitate feature extraction algorithms such as Convolution Neural Networks (CNN) \add{such as the one developed in }\cite{yuan2019}\remove{or tracking algorithms \cite{} already developed for this problem}.

\section{Acknowledgments} 
\label{sec:acknowledgments}
We would like to thank Dr. Erika Roesler (Sandia National Laboratories) for her valued time and input in this work. This paper describes objective technical results and analysis. Any subjective views or opinions that might be expressed in the paper do not necessarily represent the views of the U.S. Department of Energy or the United States Government. This work was supported by the Laboratory Directed Research and Development program at Sandia National Laboratories, a multi-mission laboratory managed and operated by National Technology and Engineering Solutions of Sandia, LLC, a wholly owned subsidiary of Honeywell International, Inc., for the U.S. Department of Energy's National Nuclear Security Administration under contract DE-NA0003525. SAND Number: SAND2020-10764 C. 

\clearpage 
\bibliographystyle{chicago}
\bibliography{cloudrefs}

\begin{thebibliography}{}

\bibitem[\protect\citeauthoryear{Berner, Bretherton, and Wood}{Berner
  et~al.}{2015}]{berner2015}
Berner, A.~H., C.~S. Bretherton, and R.~Wood (2015).
\newblock Large eddy simulation of ship tracks in the collapsed marine boundary
  layer: a case study from the monterey area ship track experiment.
\newblock {\em Atmospheric Chemistry and Physics\/}~{\em 15}, 5851--5871.

\bibitem[\protect\citeauthoryear{Blossey, Bretherton, Thornton, and
  Virts}{Blossey et~al.}{2018}]{blossey2018}
Blossey, P.~N., C.~S. Bretherton, J.~A. Thornton, and K.~S. Virts (2018,
  August).
\newblock Locally enhanced aerosols over a shipping lane produce convective
  invigoration but weak overall indirect effects in cloud-resolving
  simulations.
\newblock {\em Geophysical Research Letters\/}, 9305--9313.

\bibitem[\protect\citeauthoryear{Capaldo, Corbett, Kasibhatla, Fischbeck, and
  Pandis}{Capaldo et~al.}{1999}]{capaldo99}
Capaldo, K., J.~J. Corbett, P.~Kasibhatla, P.~Fischbeck, and S.~N. Pandis
  (1999).
\newblock Effects of ship emissions on sulphur cycling and radiative climate
  forcing over the ocean.
\newblock {\em Nature\/}~{\em 400\/}(6746), 743--746.

\bibitem[\protect\citeauthoryear{Carslaw, Lee, Reddington, Pringle, Rap,
  Forster, Mann, Spracklen, Woodhouse, Regayre, and Pierce}{Carslaw
  et~al.}{2013}]{carslaw2013}
Carslaw, K.~S., L.~A. Lee, C.~L. Reddington, K.~J. Pringle, A.~Rap, P.~M.
  Forster, G.~W. Mann, D.~V. Spracklen, M.~T. Woodhouse, L.~A. Regayre, and
  J.~R. Pierce (2013).
\newblock Large contribution of natural aerosols to uncertainty in indirect
  forcing.
\newblock {\em Nature\/}~{\em 503}, 67--80.

\bibitem[\protect\citeauthoryear{Conover}{Conover}{1966}]{conover1966}
Conover, J.~H. (1966).
\newblock Anomalous cloud lines.
\newblock {\em Journal of Atmospheric Science\/}~{\em 23}, 778--785.

\bibitem[\protect\citeauthoryear{Council}{Council}{2015}]{ci2015}
Council, N.~R. (2015).
\newblock {\em Climate Intervention: Reflecting Sunlight to Cool Earth}.
\newblock Washington, DC: The National Academies Press.

\bibitem[\protect\citeauthoryear{Eyring, Isaksen, Berntsen, Collins, Corbett,
  Endresen, Grainger, Moldanova, Schlager, and Stevenson}{Eyring
  et~al.}{2010}]{eyring10}
Eyring, V., I.~S.~A. Isaksen, T.~Berntsen, W.~J. Collins, J.~J. Corbett,
  O.~Endresen, R.~G. Grainger, J.~Moldanova, H.~Schlager, and D.~S. Stevenson
  (2010).
\newblock Transport impacts on atmosphere and climate: Shipping.
\newblock {\em Atmospheric Environment\/}~{\em 44\/}(37), 4735--4771.

\bibitem[\protect\citeauthoryear{Gunnar, Shindell, and {Br\'e on and W Collins
  and J Fuglestvedt and J Huang and D Koch and J F Lamarque and D Lee and B
  Mendoza and T Nakajima and A Robock and G Stephens and T Takemura and H
  Zhang}}{Gunnar et~al.}{2013}]{icc2013}
Gunnar, M., D.~Shindell, and F.~M. {Br\'e on and W Collins and J Fuglestvedt
  and J Huang and D Koch and J F Lamarque and D Lee and B Mendoza and T
  Nakajima and A Robock and G Stephens and T Takemura and H Zhang} (2013).
\newblock Anthropogenic and natural radiative forcing.
\newblock In T.~F. Stocker, D.~Qin, G.~K. Plattner, M.~Tignor, S.~K. Allen,
  J.~Boschung, A.~Nauels, Y.~Xia, V.~Bex, and P.~M. Midgley (Eds.), {\em
  Climate Change 2013: The Physical Science Basis. Contribution of Working
  Group I to the Fifth Assessment Report of the Intergovernmental Panel on
  Climate Change}, Chapter~8. Cambridge, United Kingdom and New York, NY, USA:
  Cambridge University Press.

\bibitem[\protect\citeauthoryear{Latham}{Latham}{1990}]{latham1990}
Latham, J. (1990).
\newblock Control of global warming?
\newblock {\em Nature\/}~{\em 347\/}(6291), 339--340.

\bibitem[\protect\citeauthoryear{Possner, H.~Wang, Wood, Caldeira, and
  Ackerman}{Possner et~al.}{2018}]{possner18}
Possner, A., H.~H.~Wang, R.~Wood, K.~Caldeira, and T.~P. Ackerman (2018).
\newblock The efficacy of aerosol--cloud radiative perturbations from
  near-surface emissions in deep open-cell stratocumuli.
\newblock {\em Atmospheric Chemistry and Physics\/}~{\em 18\/}(23),
  17475--17488.

\bibitem[\protect\citeauthoryear{Possner, Wang, Wood, and Ackerman}{Possner
  et~al.}{2018}]{possner2018}
Possner, A., H.~Wang, R.~Wood, and T.~P. Ackerman (2018).
\newblock The efficacy of aerosol-cloud radiative perturbations from
  near-surface emissions in deep open-cell stratocumuli.
\newblock {\em Atmospheric Chemistry and Physics\/}~{\em 18}, 17475--17488.

\bibitem[\protect\citeauthoryear{Riemer, West, Zaveri, and Easter}{Riemer
  et~al.}{2008}]{riemer08}
Riemer, N., M.~West, R.~Zaveri, and R.~Easter (2008, 09).
\newblock Simulating the evolution of soot mixing state with a
  particle-resolved aerosol model.
\newblock {\em Journal of Geophysical Research: Atmospheres\/}~{\em 114},
  D09202.

\bibitem[\protect\citeauthoryear{Sofiev, Sofieva, Elperin, Kleeorin,
  Rogachevskii, and Zilitinkevich}{Sofiev et~al.}{2009}]{sofiev09}
Sofiev, M., V.~Sofieva, T.~Elperin, N.~Kleeorin, I.~Rogachevskii, and
  S.~Zilitinkevich (2009, 09).
\newblock Turbulent diffusion and turbulent thermal diffusion of aerosols in
  stratified atmospheric flows.
\newblock {\em Journal of Geophysical Research: Atmospheres\/}~{\em 114},
  D18209.

\bibitem[\protect\citeauthoryear{Twomey, Howell, and Wojciechowski}{Twomey
  et~al.}{1966}]{twomey1968}
Twomey, S., H.~B. Howell, and T.~A. Wojciechowski (1966).
\newblock Comments on ``anomalous cloud lines".
\newblock {\em Journal of Atmospheric Science\/}~{\em 25}, 333--334.

\bibitem[\protect\citeauthoryear{Wang, Rasch, and Feingold}{Wang
  et~al.}{2011}]{wang2011}
Wang, H., P.~J. Rasch, and G.~Feingold (2011).
\newblock Manipulating marine stratocumulus cloud amount and albedo: a
  process-modelling study of aerosol-cloud-precipitation interactions in
  response to injection of cloud condensation nuclei.
\newblock {\em Atmospheric Chemistry and Physics\/}~{\em 11}, 4237--4249.

\bibitem[\protect\citeauthoryear{Yuan, Wang, Song, Platnick, Meyer, and
  Oreopoulos}{Yuan et~al.}{2019}]{yuan2019}
Yuan, T., C.~Wang, H.~Song, S.~Platnick, K.~Meyer, and L.~Oreopoulos (2019).
\newblock Automatically finding ship tracks to enable large-scale analysis of
  aerosol-cloud interactions.
\newblock {\em Geophysical Research Letters\/}~{\em 46\/}(13), 7726--7733.

\end{thebibliography}

\end{document}